# Spin flexoelectricity in multiferroics

Alexander P. Pyatakov[1,2*], Anatoly K. Zvezdin[1]

*1) A. M. Prokhorov General Physics Institute, 38, Vavilova st., Moscow, 119991, Russia*
*2) Physics Department, M.V. Lomonosov MSU, Leninskie gori, Moscow, 119992, Russia*

**Abstract**

Various phenomena related to inhomogeneous magnetoelectric interaction are considered. The interrelation between spatial modulation of order parameter and electric polarization, known as flexoelectric effect in liquid crystals, in the case of magnetic media appears in a form of electric polarization induced by spin modulation and vice versa. This flexomagnetoelectric interaction is also related to the effect of ferroelectric domain structure on antiferromagnetic structure, and to the magnetoelectric properties of micromagnetic structures. The influence of inhomogeneous magnetoelectric interaction on dynamic properties of multiferroics, particularly magnon spectra is also considered.

## 1. Introduction

In the last few years there was a remarkable progress in physics of multiferroics, particularly in the mechanisms of magnetoelectric coupling [1-11]. In classical review [12] that summarizes the achievement of the initial period of multiferroic era the emphasis was put on the magnetoelectric interaction biquadratic on the order parameters $F^{ME} = -\frac{1}{2}\sum_{ss'}\gamma_{ss'}^{ijkl}P^iP^jM_s^kM_{s'}^l$, where $\boldsymbol{P}$ is polarization and $\boldsymbol{M_s}$ is magnetization of sublattices (s is the number of magnetic sublattice). This term does not require any special conditions except for existence of magnetic and electric orders. Nowadays other types of interactions introduced in [12], namely the ones linear in order parameters are brought to the forefront. In particular the relation between electric polarization and existence of spatially modulated structures in the magnetic media was established [13-17], and the effects of electric field control of magnetic structure odd with respect to the electric field were revealed [18-21]. In all these cases the nonvanishing spatial derivative of magnetic order parameter $\nabla_i M_j$ creates necessary prerequisites for existence of *inhomogeneous magnetoelectric interaction*.

## 2. Inhomogeneous magnetoelectric (flexomagnetoelectric) interaction

The contribution to energy of the crystals that is linear in $\nabla_i \eta_j$ ($\boldsymbol{\eta}$ is order parameter vector) was introduced in condense matter physics long ago [22]. In context of magnetoelectric effect it appeared as *inhomogeneous magnetoelectric interaction* in theoretical papers [23,24] that was treated as the possible mechanism of long range spatially modulated spin structures. It was also introduced as a mechanism of the electric polarization generated at magnetic domain boundaries [25].

Experimentally the existence of spatially modulated spin structure (spin cycloid) has been proved in multiferroic bismuth ferrite $BiFeO_3$ [26,27]. Later it was shown that the mechanism of the spin modulation in $BiFeO_3$ is of magnetoelectric nature [14]. The origin of the spin cycloid is inhomogeneous magnetoelectric interaction whose contribution to the free energy in crystal with R3c symmetry can be written in the Lifshitz invariant-like form:

$$F_{ME} = \gamma \cdot P_z \cdot \left(L_z \cdot (\nabla \cdot L) - (L \cdot \nabla)L_z\right), \qquad (1a)$$

where **L** stands for the antiferromagnetic vector, $P_z$ is the spontaneous polarization directed along the c-axis. At temperatures below the temperature of magnetic ordering $T \ll T_N$ one can use the unit vector **n** as order parameter.

In the case of isotropic or cubic symmetry the inhomogeneous magnetoelectric interaction can be expressed in an elegant way[*] [15]:

$$F_{Flexo} = \gamma \cdot \mathbf{P} \cdot (\mathbf{n} \cdot \mathrm{div}(\mathbf{n}) + [\mathbf{n} \times \mathrm{curl}(\mathbf{n})]). \qquad (1b)$$

for the bismuth ferrite $\gamma P \sim 0.6$ erg/cm$^2$ (estimated from the wave vector of spin cycloid $\sim 10^6$ cm$^{-1}$ and its energy $\sim 6 \cdot 10^5$ erg/cm$^3$ [28]).

It is interesting to note that magnetoelectric interaction in the form (1b) is very similar to the expression used for the flexoelectric effect in nematic liquid crystal, where **n** stands for the director [15]. The electric field induces inhomogenous distribution of director in liquid crystals, and vice versa, the modulation of the director **n** induces electric polarization. This profound analogy gives grounds to name inhomogeneous magnetoelectric interaction the *flexomagnetoelectric* one, so we will use this terms as synonyms below.

The most vivid example that illustrates the relation between electric polarization **P**, wave vector **q** of the spin modulated structure and spin rotation vector **Ω** is introduced in [7]: the vectors form the vector triple $P \sim [\mathbf{q} \times \mathbf{\Omega}]$ (fig. 1).

It follows from this simple image that in magnetic ferroelectrics the spatially modulated structures is of cycloidal type (**q**⊥**Ω**), not helical one (**q**∥**Ω**). It follows also from this rule that only Neel-type domain walls can exhibit ferroelectric polarization.

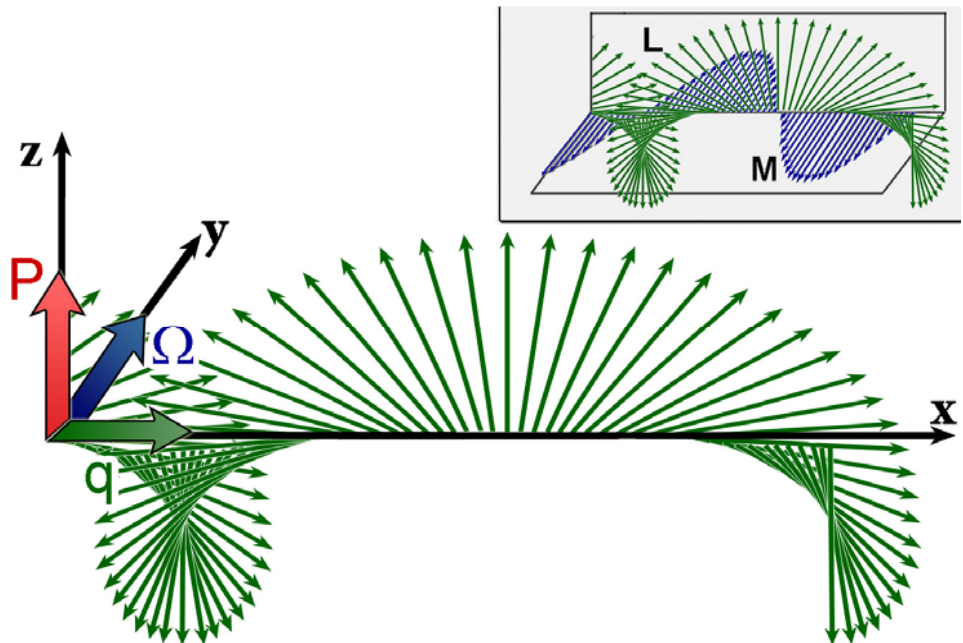

Fig. 1 a) The spin cycloid and the right-hand triple of the orthogonal vectors: the spin rotation axis *Ω*, the wave number *q*, and the polar vector *P*. In the inset: spin cycloid in vertical plane *L(x)* is associated in bismuth ferrite with the wave of magnetization *M(x)* in horizontal plane xy.

## 3. Phase transitions in bismuth ferrite BiFeO$_3$

The aforesaid spatial modulated (incommensurate) structure in multiferroic BiFeO$_3$ is the spin cycloid with a period 62 nm. The incommensurate-homogeneous (IC-H) phase transition can be considered as spontaneous nucleation of domain walls in the H-phase. The domain wall surface energy is:

$$F_{DW} = 4\sqrt{A K_u} - \pi \gamma P_s, \qquad (2)$$

---
[*] The expression (1) is given to full derivatives of order parameter.

where $P_s$ is spontaneous polarization, A and $K_u$ are constants determine the exchange stiffness and uniaxial magnetic anisotropy:

$$F_{exch} = A \sum_{i=x,y,z}(\nabla n_i)^2 = A\left((\nabla\theta)^2 + \sin^2\theta(\nabla\varphi)^2\right) \quad (3)$$

$$F_{an} = -K_u \cos^2\theta, \quad (4)$$

$\theta$ is polar angle with respect to the z-axis directed along c-axis of crystal (fig.1). The second term in domain wall energy (2) is $\int_0^\pi F_{flexo}d\theta$. Thus the condition of the phase transition $F_{DW}=0$ takes the form:

$$\gamma = \frac{4}{\pi P_s}\sqrt{AK_u}. \quad (5)$$

This expression can be derived also from the comparison of the thermodynamical potential for homogeneous and spin modulated phases, in which magnetic order parameter distribution is described by elliptic functions [27,28].

In literature another approach to the phase transition is used that is based on a harmonic approximation [13], that implies the component of **n** to depend on coordinates in the following way: $n_i(x) \sim \sin(qx+\phi_i)$. This technique allows to make the estimates with ~10% accuracy [28].

The condition (5) can be also used for estimation of the critical field $H_C$ of magnetic field induced phase transition. In this case the uniaxial anisotropy in (5) at external magnetic field H||c should be substituted by effective anisotropy $K_{eff}(H)=K_u-\chi_\perp H^2/2$, where $\chi_\perp$ is perpendicular magnetic susceptibility. In general case of arbitrary magnetic field orientation the value $H_C$ depends on the orientation with respect to the cycloid plane (see Appendix). The effective magnetic anisotropy can be also modulated by other factors: rare earth doping, magnetostriction caused by epitaxial strain in thin films etc.

The occurrence of spin cycloid in $BiFeO_3$ explains the absence of weak ferromagnetism allowed by the symmetry [5][†]. The existence of incommensurate spin structure in bismuth ferrite results in oscillating weak ferromagnetic moment that is proportional to x-component of antiferromagnetic vector *L* (fig.1 inset) so its averaged value is zero. The weak ferromagnetic moment appears only when the spin cycloid is suppressed that occurs in high magnetic fields [13, 29,30], in compounds doped by rare earth ions [31], and in thin films [32]. The later raise hopes for applications of multiferroics in spintronics [33-39].

## 4. Antiferromagnetic ordering in bismuth ferrite films with ferroelectric stripe domain structure

It is shown in numerous studies of bismuth ferrite thin films (thickness <500nm) that the incommensurate structure is suppressed [32] (see also reviews [40,41] and reference therein). That does not mean, however, that inhomogeneous magnetoelectric interaction is not relevant for the case of thin films. Even in the homogeneous antiferromagnetic state there is ferroelectric stripe domain structure [42-44] that induces modulation of antiferromagnetic vector *L* via the flexomagnetoelectric interaction, as shown in [45]. This interaction (1) manifests itself in a jump of spatial derivative $\nabla\theta$ at a boundary between the ferroelectric domains with electric polarization oriented upward «+» and downward «-»:

$$A(\nabla\theta)\big|_-^+ = \gamma P\big|_-^+ = 2\gamma P_s \quad (6)$$

---

[†] The origin of weak ferromagnetism (whether it is ferroelectrically induced or Dzyaloshinskii Moria-like) have remained disputable issue for years. See C. Fennie PRL **100**, 167203, the comment on it in PRL **102,** 249701 and the response to the comment in PRL **102,** 249702.

The graphical image of antiferromagnetic vector modulation $\theta(x)$ is shown in figure 2. Thus, in presence of ferroelectric domain structure the ground state of antiferromagnetic system is not uniform anymore. If antiferromagnetic and ferroelectric domain structures coexist, then the above-mentioned effect will manifest as a pinning of antiferromagnetic domain structure at ferroelectric domain walls. Such a coupling of ferroelectric and antiferromagnetic domains, particularly in manganites, was reported in [46].

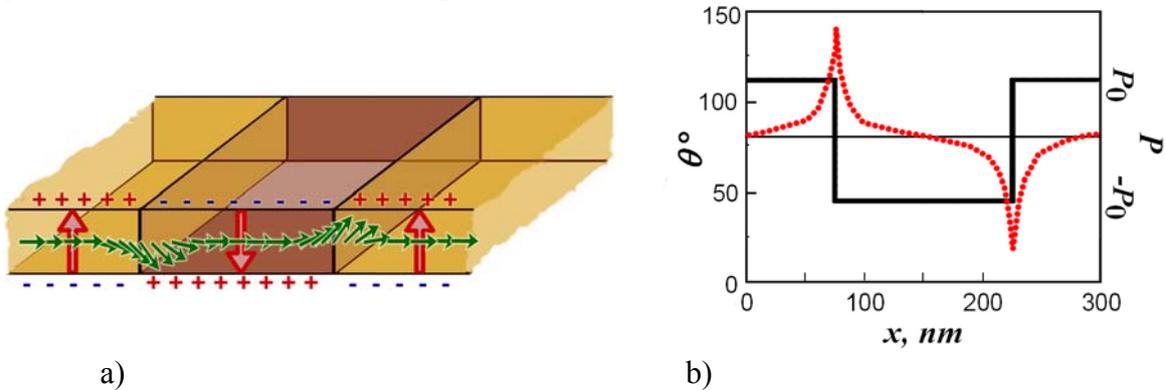

a)                                                                b)

Fig. 2 The modulation of the antiferromagnetic vector orientation induced by ferroelectric structure in multiferroics with flexomagnetoelectric interaction: a) schematic view (solid arrows exemplifies antiferromagnetic vector, open arrows are the electric polarization vectors) b) the dependence of the polar angle on coordinate $\theta(x)$ (dotted line) superimposed on the ferroelectric structure (solid line) [45].

## 5. The influence of flexomagnetoelectric interaction on magnon spectra in multiferroics

The existence of flexomagnetoelectric interaction in the homogeneous state, though in a latent form, manifest itself not only in cycloid and the aforementioned static structures in films but also in dynamical properties of multiferroics, particularly in magnon (electromagnon [47]) spectra. The influence of the flexomagnetoelectric interaction on magnon spectra in materials with incommensurate structures was considered in [48-51]. In thin films spin modulated structure is usually absent and the magnetoelectric interaction (1) seems to be irrelevant here (see [52], for example). However, it was shown in [53; 54] that the flexomagnetoelectric interaction (1) influences considerably on magnon spectra even in the homogeneous state: the dispersion curve has minimum at finite wave vectors for waves propagating in the direction normal to the (P,L) plane in the case of multiferroic [54] or perpendicular to the external electric field in the case of centrosymmetric ferromagnetic media [53].

Thus in thin films of bismuth ferrite inhomogeneous magnetoelectric interaction causes the coupling between modes of spin wave propagating along weak ferromagnetic moment $k \| M_0$. This coupling appears as a minimum of dispersion curve w(q) at the wave vector of the spin cycloid $q_0$ (fig. 3). For $BiFeO_3$ $q_0$ is ~$10^6$ cm$^{-1}$. Another interesting feature of magnons in $BiFeO_3$ consists in nonreciprocity of magnon propagation along and opposite to antiferromagnetic vector that is parallel with toroidal moment [54]. These peculiarities can appear in Raman and Brillouin light scattering that can be used to obtain the parameters of magnetoelectric interactions.

Of special interest are practical aspects, as magnons in multiferroics can be excited and controlled by electric field. The promising design of ultrasmall logic devices based on electromagnon excitations was proposed [55,56]. The conventional semiconductor technology at the present level of miniaturization encounters the problem of large electric fields approaching the threshold one. The alternative approach based on the spin transfer effects requires large current densities (~$10^6$ A/cm$^2$ or even more). The implementation of logic devices based on long living magnons in magnetic dielectrics might be a solution of this problem.

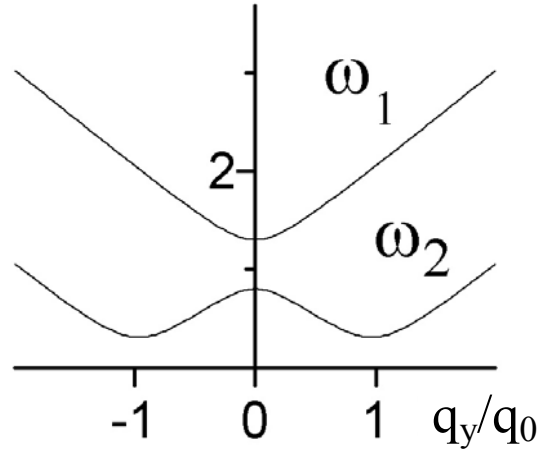

Fig. 3 The minimum in the dispersion curve for magnon in multiferroic with flexomagnetoelectric interaction in homogeneous magnetic state. $q_0$ is the wave vector of spin cycloid in volume crystal. [54]

## 6. Surface flexomagnetoelectric effect

Special conditions realized in thin films of bismuth ferrite lead to suppression of spatially modulated structures that are observed in volume crystals. However the inverse situation is not only possible but is even more natural: the existence of center of symmetry in the symmetry group of crystal forbids flexomagnetoelectric interaction while at the surface the inversion symmetry is broken and the symmetry restriction is lifted.

The possibility of spontaneous spin modulated structure occurrence at the surface of centrosymmetric magnetic crystal and in thin films was predicted in [57], the condition of phase transition between homogeneous and incommensurate phase was formulated (for details see Appendix B).

This effect was discovered in Mn monolayers [58] and in two monolayers of Fe epitaxially grown on (110) tungsten substrate [59]. Using polarized electron scanning tunneling microscope they observed magnetic modulation with a period ~0.5 nm [58]. Measurements with probes of different magnetic moment orientation enabled to determine that the structure in monolayer corresponds to the spin cycloid [58], while in two monolayers of Fe it was the domain wall of Neel type with certain chirality [59].

## 7. Flexomagnetoelectric effect as the origin of improper ferroelectricity in multiferroics

As was mentioned above, the central symmetry breaking leads to the formation of spatially modulated structures. However the inverse effect is possible: spatially modulated structure lowers the symmetry of the crystal and ferroelectric polarization appears.

This mechanism is supposed to cause electric polarization in orthorhombic manganites, $RMnO_3$ (R=Tb, Dy, Gd) [9,16-19], in vanadate $Ni_3V_2O_8$ [60], and in hexaferrite $Ba_2Mg_2Fe_{12}O_{22}$ [17]. In multiferroic $MnWO_4$ the ferroelectric domains formed by spin spiral with opposite chiralities was observed [61]. The magnetoelectric effects such as magnetic control of electric polarization [16,17] as well as transformation of spatially modulated structure under influence of electric field [18,19] are also attributed to the flexomagnetoelectric interaction.

Indeed, it follows from the figure 1 that the reversal of electric polarization should lead to the reorientation of the rotation vector $\mathbf{\Omega}$ with respect to the wave vector $\mathbf{q}$, i.e. to the reversal of the chirality. This effect was demonstrated in [18, 19].

In the case of ferroelectrics such as BiFeO$_3$ and BaMnF$_4$, in which large spontaneous electric polarization is present the additional polarization induced by spin modulation[‡] can be detected only under special circumstances, namely at the IC-H phase transitions expressed by equation (4).

Flexomagnetoelectric polarization can be found from the contribution to the thermodynamic potential:

$$\Delta P = -\frac{\partial F_{Flexo}}{\partial E} = \gamma \kappa \frac{d\theta}{dx}, \qquad (6)$$

where $\kappa$ is electric susceptibility: $P=\kappa E$, $E$ is electric field.
The polarization averaged over the period of cycloid is:

$$\langle \Delta P \rangle = \frac{1}{\lambda}\int_0^{2\pi} \Delta P(x)\frac{dx}{d\theta} d\theta = \frac{2\pi}{\lambda}\gamma\kappa, \qquad (7)$$

where $\lambda$ is the period of the cycloid.

In figure 4 the magnetoelectric anomalies are shown that appear as jumps of electric polarization in certain critical field. These anomalies can be explained as additional polarization that is related to the spin cycloid.

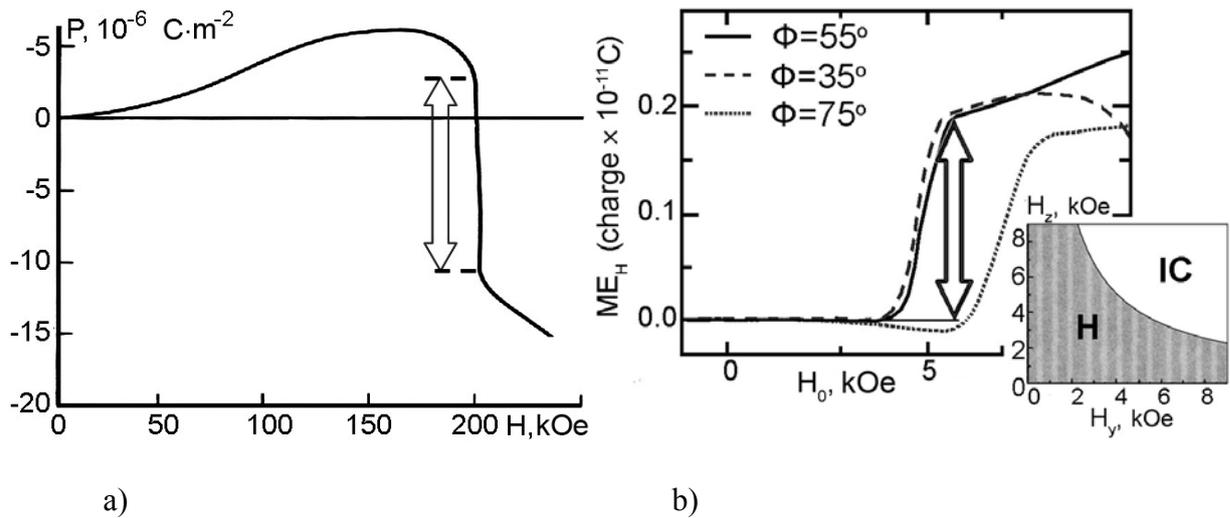

a)                                         b)
Fig. 4 The magnetoelectric anomalies at phase transitions: a) incommensurate phase-homogenous state in BiFeO$_3$ [13] б) homogenous state – incommensurate state in BaMnF$_4$ [62] (in the inset there is the phase diagram in the coordinates ($H_y$, $H_z$), **H** is homogenous state, **IC** is incommensurate phase).

In figure 4 a) the magnetoelectric curve for the BiFeO$_3$ is shown. The jump of polarization is observed in the field ~200kOe when cycloid is suppressed.

Interesting situation occurs in magnetoelectric BaMnF$_4$. The jump of electric polarization was observed in the magnetic field oriented in bc-plane when the angle of magnetic field with respect to the b-axis is in the vicinity of 45° (fig.4b). This effect occurs in $H_C \sim$ 5kOe and still lacking theoretical interpretation. We believe this effect is a flexomagnetoelectric one.

Indeed the symmetry of BaMnF$_4$ (class 2, space group A2$_1$am) allows inhomogeneous magnetoelectric interaction:

$$F_{Flexo} = -\left(\gamma_{01} + \gamma_{11}H_y H_z\right)P_x \frac{\partial \theta}{\partial x} = -\left(\gamma_{01} + \frac{\gamma_{11}H^2}{2}\sin 2\Phi\right)P_x \frac{\partial \theta}{\partial x} \qquad (8)$$

---

[‡] In general case the calculation of electric polarization in modulated magnetics is a self-consistent problem. However in the case of proper ferroelectric $T_C \gg T_N$ the electric polarization can be considered as a sum of spontaneous polarization $P_S$ and perturbation $\Delta P$ caused by spin modulation.

where Φ is the angle of the magnetic field with respect to the b-axis of the crystal.
It differs from the conventional form (1) in the way that the magnetoelectric coefficient γ depend on the value and orientation of the magnetic field. One can easily find from (8) that magnetoelectric coefficient is maximum at the angle Φ=45˚, so the critical magnetic field at this orientation is minimum and is equal to 4.5kOe (fig. 4b).

## 8. Thin films of iron garnets and flexomagnetoelectric effect at room temperatures

If one compares the number of papers devoted to bismuth ferrite with the reports on other room temperature magnetoelectric the striking disproportion of research activity will become evident. For example the potential of iron garnets films seems to be underestimated as the linear magnetoelectric effect measured by electric field induced Faraday rotation in this material is an order larger than in classical magnetoelectric $Cr_2O_3$ [63]. Iron garnet films are probably the most convenient object to study micromagnetism, so the relation between the micromagnetic structure and flexomagnetoelectric interaction can be illustrated in this material.

The influence of electric field on micromagnetic structure was predicted theoretically in the series of works [7, 15, 25, 64, 65]. In this context the magnetic domain walls [25,65], magnetic vortices [7] and vertical Bloch line [64] were considered and it has been shown that various electric charge distributions are associated with them. In the recent paper [65] the nucleation of the Neel-type domain wall in magnetic media at critical electric field has been predicted. This phenomenon, by no means, is interesting as prototype electric write/magnetic read memory.

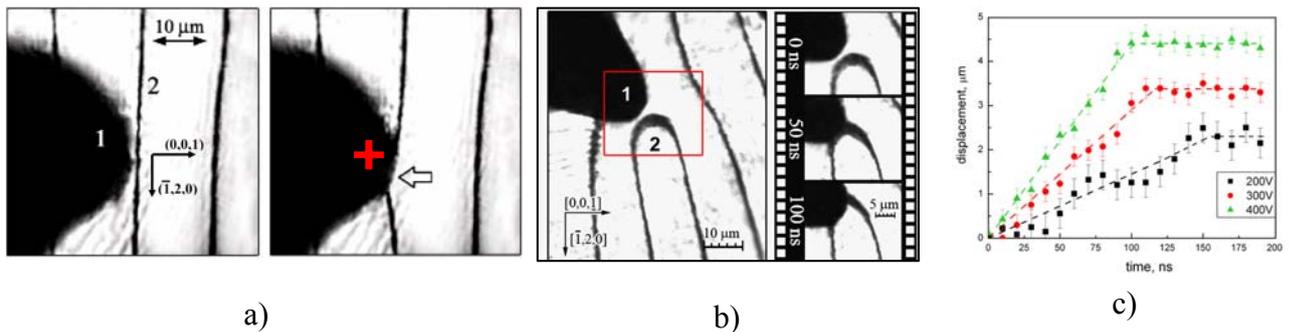

a)  b)  c)

Fig. 5. Electric field control of the domain wall in epitaxial iron garnet films: a) the magnetooptical image of iron garnet film in the transmitted light: dark lines are the magnetic domain boundaries, 1 is the tip electrode, 2 is a domain wall [20].
b) The stripe domain head motion induced by electric field [21] c) the time dependence of the displacement of the domain boundary at different potential at the electrode [21].

Although the electric field induced nucleation of Neel domain wall was not confirmed experimentally so far, the motion of domain walls in magnetic material under influence of electric field has been observed in the 10μm thickness epitaxial iron garnet films, grown on the GdGG substrate [20; 21]. In figure 5 the effect of electric field from the tip electrode is shown. The tip electrode potential positive with respect to the substrate causes the attraction of the domain wall to the tip (fig. 5 a), and the negative one results in repulsion. When the voltage is switched off the domain wall like a stretched string goes to the initial equilibrium state. Not always the transformation of micromagnetic structure had a reverse character: if the modified state were more stable than the initial one the domain walls would remain in the new position.
The basic features of the effect are:
(i) The change of the sign of the effect at electric polarity reversal;
(ii) The independence of the effect from the magnetic polarity of the domain over which the tip was located (T-even effect);
(iii) the crucial role of anisotropy of the films (effect was observed in films with (210) and (110) substrate orientations and was not observed in (111) films).
These properties evidences for its flexomagnetoelectric nature.

Indeed, the dependence on the electric polarity and T-evenness follow directly from the equation (1), while the dependence on the substrate orientation is related to the difference between the micromagnetic configuration in highly symmetrical (111) films and ones in the films with low symmetry (110), (210) substrates. In the (111) films the direction of the anisotropy axis is along the normal to the film, the boundaries between domains are of Bloch type ($div\mathbf{M} = 0$; $[\mathbf{M} \times curl\mathbf{M}] = 0$, provided that |$\mathbf{M}$|=const) and therefore the effect is not observed. At the same time in the (210) and (110) films the anisotropy axis is inclined to the normal that causes the Neel component in domain walls so the flexomagnetoelectric interaction is nonvanishing [21].

The possibility of the domain wall motion in electric field is also discussed in [65], and it is shown that the domain wall velocity should be proportional to the electric field gradient. The experimental studies of domain wall dynamics in pulse electric field (fig. 5 b, c) show that the velocity increases with increasing tip electrode potential. The ultimate displacement of the wall demonstrates similar dependence on the voltage.

Comparison of the results of measurements in pulse electric field with their analogues in pulse magnetic field gives us the quantitative measure of the effect: the voltage 500 V (i.e. the electric field 1 MV/cm) creates the effect similar to the one of 50 Oe magnetic field [21]. From this data we find the threshold field of single domain wall nucleation $E_t = 4\sqrt{KA}/\pi\gamma$ ~200 MV/cm. This value is too large to be obtained at normal circumstances.

More practicable in this context looks the novel concept of magnetic storage based on domain wall shift [66]. At the size of technological junction ~100 nm and the domain wall velocity 100 m /s the switching time of the element might be ~1 ns.

There are other interesting manifestation of flexomagnetoelectric interaction in Nuclear Magnetic Resonance [67], Electron Spin Resonance [68] and probably in Second Harmonic Generation [69,70] but they are beyond the scope of this paper.

Authors are grateful to A.M. Kadomtseva, Yu.F. Popov, G.P. Vorob'ev, A.A. Mukhin, Z.V. Gareeva, A.S. Logginov, A.V. Nikolaev, H. Schmid, D.I. Khomskii and M. Bibes for collaboration and valuable discussions. The support of RFBR № 08-02-01068-a and Progetto Lagrange-Fondazione CRT is acknowledged.

## Appendix A: Magnetic field induced phase transition

Let us find the dependence of critical magnetic field on the orientation with respect to the cycloid plane in harmonical approximation. That means anisotropy does not influence on the shape of the cycloid and the dependence of polar angle is linear in coordinate:

$$\theta = qx \qquad (A.1)$$

The wave vector of the spin cycloid $q_0$ that correspond to the minimum of energy can be found from the following condition:

$$\frac{\partial(F_{exch} + F_{flexo})}{\partial q} = \frac{\partial(Aq^2 - \gamma \cdot P_z q)}{\partial q} = 0, \qquad (A.2)$$

therefore:

$$q_0 = \frac{\gamma \cdot P_z}{2A}. \qquad (A.3)$$

In external magnetic field $H=(0, H_y, H_z)$ the contribution to the free energy in cycloidal state will be:

$$F_{cycloid} = Aq_0^2 - \gamma \cdot P_z q_0 - K_u \cos^2\theta - \frac{\chi_\perp H_z^2}{2}\sin^2\theta - \frac{\chi_\perp H_y^2}{2}, \qquad (A.4)$$

using (A.3) we obtain for the value of energy averaged over the period $\lambda$ of the cycloid:

$$\langle F_{cycloid} \rangle_\lambda = -Aq_0^2 - \frac{K_u}{2} - \frac{\chi_\perp H_z^2}{4} - \frac{\chi_\perp H_y^2}{2}. \qquad (A.5)$$

For homogeneous state at which antiferromagnetic vector is perpendicular to the magnetic field and the c-axis ($\theta=90$, $\varphi=0$) the free energy is:

$$F_{hom} = -m_s H_y - \frac{\chi_\perp H_z^2}{2} - \frac{\chi_\perp H_y^2}{2}, \qquad (A.6)$$

where $m_s$ the weak ferromagnetic moment.

The critical field of phase transition $H_C$ can be found from the equation (A.5) and (A.6):

$$H_C = \frac{-2m_s \sin\psi + 2\sqrt{m_s^2 \sin^2\psi + \chi_\perp \cos^2\psi(Aq_0^2 + K_u/2)}}{\chi_\perp \cos^2\psi} \qquad (A.7)$$

where $\psi$ is the angle of magnetic field with respect to the c-axis: $H_C=(0, H_C \sin\psi, H_C \cos\psi)$.

For BiFeO$_3$ $A=3\cdot10^{-7}$ erg/cm, $q_0= 10^6$cm$^{-1}$ (corresponds to $\lambda=62$ nm), $\chi_\perp = 4.7\cdot10^{-5}$, $m_s\sim 5$G [45] and $K_u = 3\cdot10^5 \frac{erg}{cm^3}$ (that include magnetic anisotropy $K_u^0 = 6\cdot10^5 \frac{erg}{cm^3}$, and the contribution of weak ferromagnetism $K_{DM} = -\frac{m_s^2}{2\chi_\perp} \approx -3\cdot10^5 \frac{erg}{cm^3}$ [68]). The calculated dependence is shown in figure 6. It should be stressed that these results are obtained neglecting the deformation of the cycloid shape in the external field, i.e. the cycloid remained harmonical described by equation A.1 and the wave vector $q_0$ (A.3).

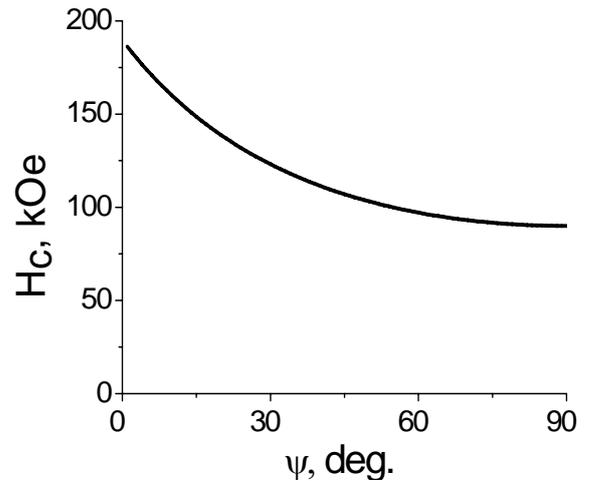

Fig. 6 The orientational dependence of critical field of phase transition from spin modulated to homogeneous field. The $\psi$ is the angle of magnetic field with respect to the c-axis.

## Appendix B: Spatially modulated structures in thin films

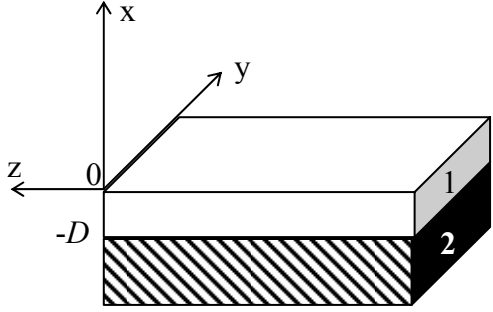

Fig.7 Geometry of the problem: (1) is a film of thickness D, (2) is isotropic substrate.

In this appendix we analyze the possibility of the formation in thin films the spatially modulated structure induced not by volume interaction but by the presence of a Lifshitz type invariant in the surface energy. Let us consider a film medium with the orientational order parameter **n** ($|\mathbf{n}|=1$). In the case of a ferromagnet this is a unit vector directed along a local magnetization, while in the case of antiferromagnetic media it is vector of antiferromagnetism. The geometry of the film deposited onto an isotropic substrate and a Cartesian coordinate system are presented in Fig.7.

Let us present the energy density of the film as
$$F = A\left((\nabla\theta)^2 + \sin^2\theta(\nabla\varphi)^2\right) + \beta(x)I - (K_1 + K_s)n_x^2, \qquad (B.1)$$
where the first summand is the exchange energy, the second summand is a Lifshitz-like invariant
$$I = n_z\frac{\partial n_x}{\partial z} - n_x\frac{\partial n_z}{\partial z} - n_x\frac{\partial n_y}{\partial y} + n_y\frac{\partial n_x}{\partial y}, \qquad (B.2)$$
and the third summand is the energy density of the anisotropy involving the general case of the bulk energy $K_1$ and the near surface energy $K_s(x)$. In magnetic films $K_s$ is the Neel magnetic surface anisotropy. The coefficients $\beta(x)$ and $K_S(x)$ are nonzero in only narrow intervals near the upper and lower film surface. When integrated over x they give the "real" surface coefficients $\beta_S$ and $K_S$. The polar $\theta$ and azimuthal $\varphi$ angles of the director are determined in the usual manner in the coordinated system plotted in Fig. 7, the angle $\theta$ being counted from the z-axis and the angle $\varphi$ from the x-axis. In angular variables the second summand assumes the form
$$\beta(x)I = \beta(x)\left[\nabla_z\theta\cos\varphi - \frac{\sin 2\theta}{2}\sin\varphi\cdot\nabla_z\varphi - \sin^2\theta\cdot\nabla_y\varphi\right]. \qquad (B.3)$$

We shall treat the anisotropy energy as a disturbance, the energy density $F^0$ and the energy $E^0$ integrated over the volume of the sample as being equal to
$$F^0 = A\left((\nabla\theta)^2 + \sin^2\theta(\nabla\varphi)^2\right) + \beta(x)I \qquad (B.4)$$
$$E^0 = \int F^0 dxdydz. \qquad (B.5)$$

The Euler-Lagrange equation for the functional (B.5) have the form
$$-2A(\nabla\theta)^2 - \beta\nabla_z(\cos\varphi) + 2S\sin\theta\cos\theta(\nabla\varphi)^2 + \beta\left(-\cos 2\theta\sin\varphi\cdot\nabla_z\varphi - 2\sin\theta\cos\theta\cdot\nabla_y\varphi\right) = 0 \qquad (B.6)$$
$$-2A\nabla\sin^2\theta(\nabla\varphi)^2 + \beta\nabla_z(\sin\theta\cos\theta\sin\varphi) + \beta\nabla_y(\sin^2\theta) + \beta\left(-\sin\varphi\cdot\nabla_z\theta - \frac{\sin 2\theta}{2}\cos\varphi\cdot\nabla_z\varphi\right) = 0. \qquad (B.7)$$

The Euler-Lagrange equation (B.6) and (B.7) obviously have trivial solution $\theta=\pi/2$, $\varphi=0$ (K>0) and $\theta=\pi/2$, $\varphi=\pi/2$ (K<0). In the case of spatially homogeneous structure, the total film energy with allowance for anisotropy is equal to
$$E^{hom} = -(K_S + K_1 D). \qquad (B.8)$$

We shall now consider the inhomogeneous states. One can readily see that equation (B.6) has solution $\theta=\pi/2$. Substitution of $\theta=\pi/2$ into (B.7) brings equation (B.7) to the form
$$\nabla^2\varphi = 0. \qquad (B.9)$$

Without loss of generality, the solution of equation (B.9) can be represented as $\varphi=\varphi(y)$. So, equation (B.6),(B.7) have the following particular solution

$$\theta=\pi/2;\ \varphi=qy. \quad (B.10)$$

This solution describes the spatially modulated (cycloidal) structure and q is its vector. To find the constant q, we shall substitute (B.10) into (B.4) to obtain

$$E^0 = Aq^2 DL_z L_y - q\int_{-D}^{0}\beta(x)dx\iint dydz = Aq^2 DL_y L_z - q\beta_s L_y L_z, \quad (B.11)$$

where $\beta_s = \int_{-D}^{0}\beta(x)dx$ and $L_y L_z$ is the area over which the integration was taken.

Minimization of the energy (B.11) yields

$$q = \frac{\beta_s}{2AD} \quad (B.12)$$

from which the structure period λ is equal to

$$\lambda = \frac{2\pi}{q} = \frac{4\pi AD}{\beta_s} \quad (B.13)$$

Including the energy of anisotropy by thermodynamic perturbation theory

$$E = E^0 + \langle F_{an}\rangle,$$

where $F_{an} = (K_1 + K_s)n_x^2$ and $\langle ...\rangle$ means averaging over the film volume if the angle $\theta$ and $\varphi$ are determined from equation in zero approximation (in this case from equations (B.6), (B.7)), we arrive at the total film energy (per unit area):

$$\sigma = \frac{1}{L_z L_y}E = Aq^2 D - q\beta_s + \frac{\int F_{an}dV}{L_y L_z} = -\frac{\beta_s^2}{8AD} - \left(\frac{K_s}{2} + \frac{K_1 D}{2}\right). \quad (B.14)$$

Comparing the energy (B.14) of the spatially modulated (cycloidal) structure with the energy (B.8) of the homogenous structure, one can establish the phase existence regions for this structure. The interfaces are defined by the inequalities

$$-\frac{\beta_s^2}{8AD} - \left(\frac{K_s}{2} + \frac{K_1 D}{2}\right) \le -(K_s + K_1 D), \quad (B.15)$$

It is convenient to represent the inequalities (B.15) in the form

$$4\pi^2 A \ge \lambda^2\left(\frac{K_s}{D} + K_1\right) \quad (B.16)$$

Where all the parameters involved in the inequality can be found experimentally.